\documentclass[pra,showpacs]{revtex4}
\usepackage{makeidx}
\usepackage{amsmath}
\usepackage{graphicx}
\usepackage{epsfig}
\usepackage[usenames]{color}

\newcommand{\beq}{\begin{equation}}
\newcommand{\eeq}{\end{equation}}
\newcommand{\beqa}{\begin{eqnarray}}
\newcommand{\eeqa}{\end{eqnarray}}
\newcommand{\ba}{\begin{array}}
\newcommand{\ea}{\end{array}}

\begin{document}

\title{Collapse of triaxial bright solitons in atomic Bose-Einstein 
condensates} 
\author{G. Mazzarella$^{1}$ and L. Salasnich$^{1,2}$}
\affiliation{$^1$Dipartimento di Fisica ``Galileo Galilei'', 
Universit\`a di Padova, Via Marzolo 8, 35131 Padova, Italy \\
$^{2}$CNR-INFM and CNISM, Unit\`a di Padova, 
Via Marzolo 8, 35131 Padova, Italy}

\begin{abstract}
We study triaxial bright solitons made of attractive Bose-condensed atoms 
characterized by the absence of confinement 
in the longitudinal axial direction but trapped 
by an anisotropic harmonic potential in the transverse plane. 
By numerically solving the three-dimensional 
Gross-Pitaevskii equation we investigate the effect of the transverse 
trap anisotropy on the critical interaction strength above 
which there is the collapse of the condensate. The comparison 
with previous predictions [Phys. Rev. A {\bf 66}, 043619 (2002)] shows 
significant differences for large anisotropies. 
\end{abstract}


\maketitle

The experimental achievement of quantum degeneracy with 
ultracold alkali-metal atoms \cite{Anderson,Inguscio} has opened 
the possibility of studying various topological configurations of the 
Bose-Einstein condensate (BEC) with repulsive or attractive 
inter-atomic interaction \cite{Drazin}. Dark 
solitons in repulsive BECs have been experimentally achieved ten 
years ago \cite{Burger}, while bright solitons have been 
detected only more recently in two different experiments 
\cite{Khay,Stre} involving attractive BECs of $^{7}$Li vapors. In 
these latter experiments, an optical red-detuned laser beam 
generated along the axial direction of the sample is used to trap 
the attractive BEC by a cylindric isotropic transverse 
confinement; the BEC propagates along the longitudinal axis of the 
cylinder without relevant spreadings. Recently, also $^{85}$Rb 
atoms have been used to achieve the Bose-Einstein condensation and 
investigate the formation of bright matter-wave solitons during 
the collapse \cite{wieman}. 

Many theoretical works have been devoted to the study of 
cigar-shaped and axially symmetric bright-soliton configurations, 
also in presence of an axial periodic potential 
\cite{3,4,5,gammal,6,7,8,9,10,11}. The 
transverse confinement produced by the isotropic harmonic
potential in the cylindric radial direction plays a crucial role 
in giving rise to single \cite{3,4,5,gammal,6} or multiple \cite{7,8} 
metastable bright solitons, which collapse above a critical number
of particles \cite{5,gammal,6,8}. These theoretical investigations showed
that increasing the inter-atomic strength, e.g. by Fano-Feshbach
resonances, makes the bright soliton less cigar-shaped. In 
particular, a quasi-spherical shape is achieved only when the
interaction strength approaches the critical value that signs the 
collapse \cite{5,6}.

In this paper we study an attractive BEC trapped by 
an anisotropic harmonic potential in the transverse plane 
and without confinement in the axial direction. 
Under this trapping condition the BEC admits 
stable bright-soliton configurations, which are are generally triaxial. 
The deformation of the transverse anisotropic confinement can 
be described by two independent parameters 
\cite{SalaLP} or by a unique quantity, the ellipticity \cite{Martin}. 
In both cases, by using 
the numerical integration of the three-dimensional Gross-Pitaevskii equation,  
we investigate as a function of the transverse trap anisotropy  
the critical interaction strength above which the triaxial 
soliton collapses, i.e. it shrinks to the zero-size ground-state 
of infinite negative energy. We compare our results with previous 
numerical predictions \cite{gammal} and find that our 
stability domain is significantly smaller. 

Let us consider an attractive BEC without confinement in the  
axial direction $z$ and confined in the transverse plane $(x,y)$ 
by the anisotropic harmonic potential 
\beq 
U(x,y) = {m\over 2} \left( \omega_1^2 x^2
+ \omega_2^2y^2 \right) \; , 
\label{harmonicpotential} 
\eeq 
where $m$ is the mass of a Bose-condensed atom, and $\omega_1$,
$\omega_2$ are the two frequencies of the harmonic confinement.
With the aim of working in scaled units we set 
\beq 
\omega_1 = 
\lambda_1 \ \omega_{\bot}  \; , \quad\quad \omega_2 = \lambda_2 \
\omega_{\bot} \; . 
\eeq 
In particular, if $a_{\bot}=(\hbar/(m\omega_{\bot}))^{1/2}$ 
is used as characteristic harmonic length of the system, then lengths may be
measured in units of $a_{\bot}$ and energies in units of $\hbar
\omega_{\bot }$. 

The dynamics of an attractive BEC can be accurately described 
by the adimensional time-dependent 3D Gross-Pitaevskii equation (3D GPE), 
given by 
\beq 
i\hbar {\partial \over \partial t} \Psi = 
\left[ -{\frac{1}{2}}\nabla^{2} + {1\over 2} \left(
\lambda_1^2 x^2 +\lambda_2^2 y^2 \right) - 2\pi g |\Psi
|^{2}\right] \Psi \; , 
\label{tdgpe}
\eeq 
where $\Psi({\bf r},t)$ is the macroscopic wave function of the condensate
and 
\beq
g= {2 N |a_s| \over a_{\bot} }
\eeq
is the interaction strength, with $N$ the number of atoms 
and $a_s<0$ the $s$-wave 
scattering length of the inter-atomic potential. 
Setting 
\beq 
\Psi({\bf r},t) = \psi({\bf r}) \ e^{-i\mu t} \; ,  
\eeq
from Eq. (\ref{tdgpe}) one finds the stationary 3D GPE  
\beq 
\left[ -{\frac{1}{2}}\nabla^{2} + {1\over 2} \left(
\lambda_1^2 x^2 +\lambda_2^2 y^2 \right) - 2\pi g |\psi
|^{2}\right] \psi = \mu \ \psi \; , 
\label{3dgpe} 
\eeq 
where the chemical potential $\mu$ is fixed by the normalization 
\beq 
\int |\psi({\bf r})|^2 \ d^3{\bf r} = 1 \; . 
\label{norma} 
\eeq 
Stable solutions of Eq. (\ref{3dgpe}) are called bright solitons 
\cite{3,4,5,gammal,6}. They correspond to an attractive BEC 
with a self-confinement along the $z$ axis. 

\begin{figure}
\centering
\includegraphics[width=8cm,clip]{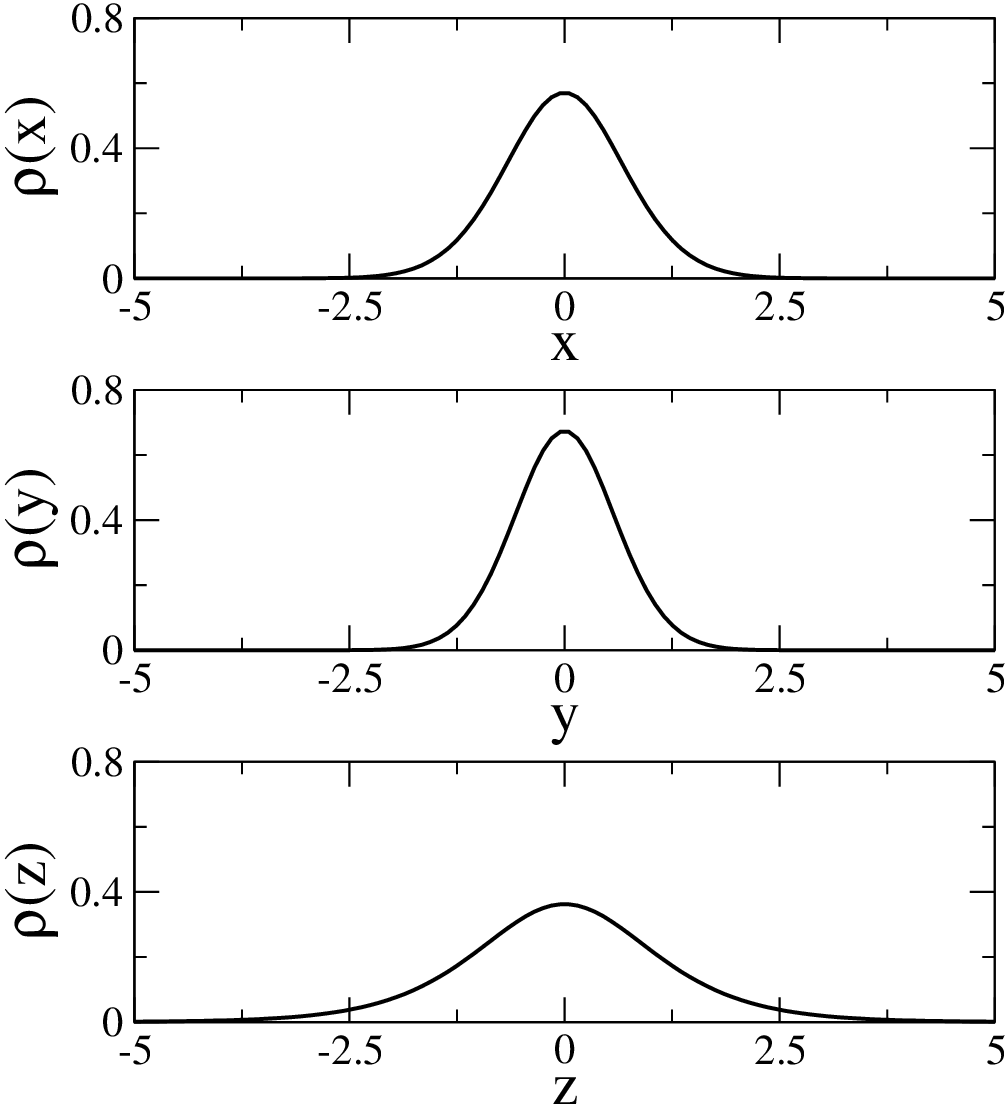}
\caption{Integrated density profiles $\rho(x)$,
$\rho(y)$ and $\rho(z)$ of the triaxial bright soliton along the three
Cartesian axes obtained by the numerical integration of the 3D GPE. 
Interaction strength $g=2N|a_s|/a_{\bot}=1.2$ and 
trap ellipticity $\epsilon=0.4$.} 
\label{fig1}
\end{figure}

To determine the solutions of Eq. (\ref{3dgpe}) 
we solve Eq. (\ref{tdgpe}) by using a finite-difference 
Crank-Nicolson algorithm with imaginary time \cite{sala-numerics} 
and a spatial mesh of $160\times 160 \times 160$ points 
(for details see the Appendix). 
In this way we determine the wave function $\psi({\bf r})$ of the 
metastable bright soliton 
and we can also calculate the integrated density profiles 
of the triaxial bright soliton along the three spatial directions, 
given by 
\beqa 
\rho(x) = \int |\psi({\bf r})|^2\ dy \ dz \; , 
\\
\rho(y) = \int |\psi({\bf r})|^2\ dx \ dz \; , 
\\
\rho(z) = \int |\psi({\bf r})|^2\ dx \ dy \; . 
\eeqa 

To study the critical strength above which there is the collapse, 
we consider first the simpler case of elliptic transverse confinement. 
As explained by Jamaludin {\it et al.} \cite{Martin}, 
it is possible to consider an elliptic transverse harmonic confinement 
and consequently to parametrize 
the transverse anisotropy of the harmonic 
confining potential (\ref{harmonicpotential}) by using a unique parameter, 
the trap ellipticity $\epsilon$. In terms of the 
ellipticity $\epsilon$, the scaled harmonic frequencies are written as 
\beq 
\lambda_1 = \sqrt{1 - \epsilon} \; , \quad\quad \lambda_2
= \sqrt{1 + \epsilon} \; , 
\label{ellisse} 
\eeq 
with $\epsilon$ restricted to the interval $[-1,1]$. 
Clearly $\epsilon=0$ corresponds to the isotropic transverse
confinement, while $\epsilon=\pm 1$ implies the absence of
confinement along the $x$ axis ($\epsilon=1$) or along the $y$
axis ($\epsilon=-1$). 

\begin{figure}
\centering
\includegraphics[width=8.2cm,clip]{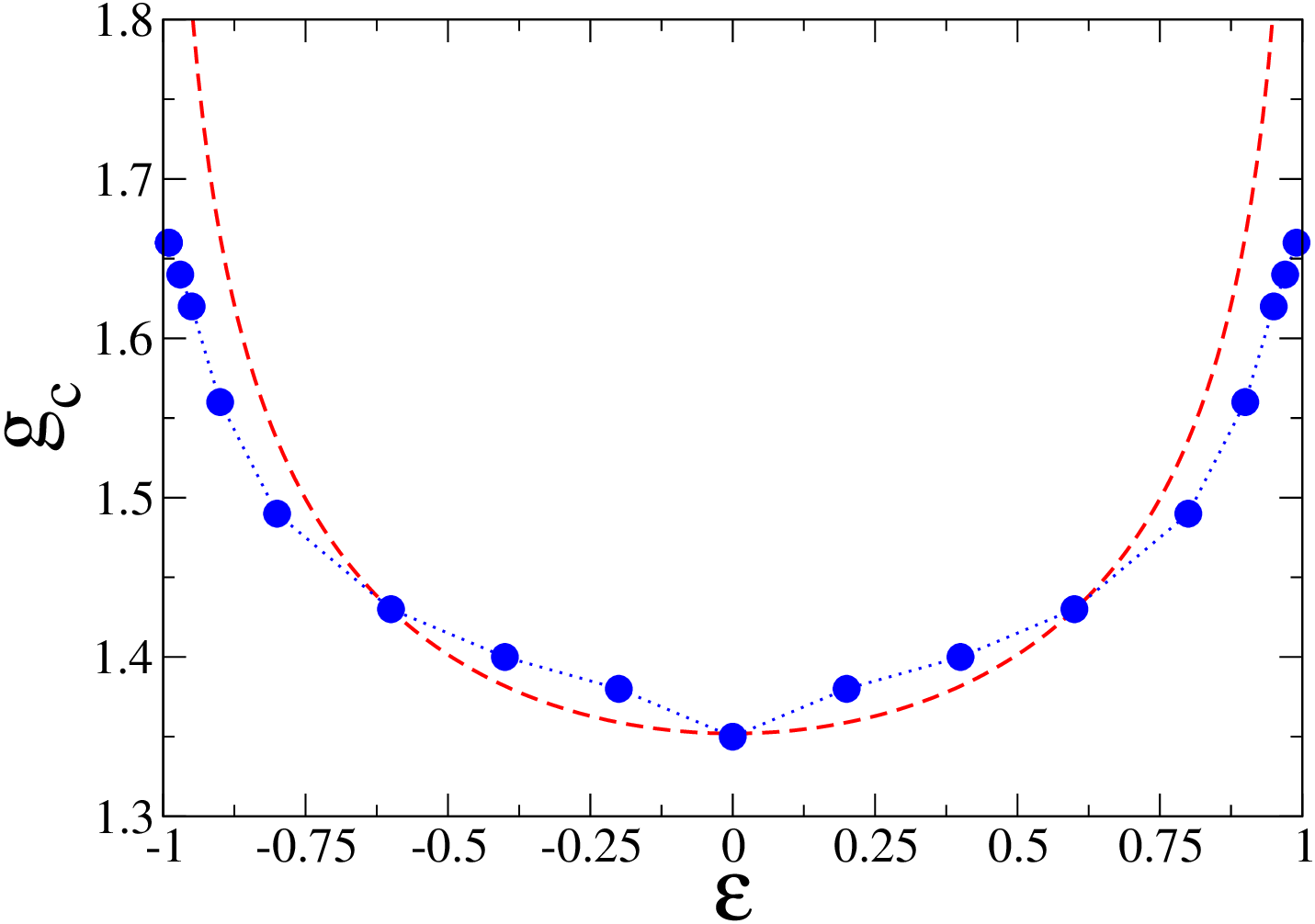}
\caption{Critical strength $g_c$ for
the collapse of the triaxial bright soliton 
as a function of the ellipticity $\epsilon$ of the elliptic 
transverse harmonic confinement, Eq. (\ref{ellisse}). 
Filled circles: numerical results 
obtained with the 3D GPE. Dashed line: prediction 
of Eq. (\ref{brazil}).}
\label{fig2}
\end{figure}

As an example, in Fig. \ref{fig1} we plot the density profiles 
of the bright soliton choosing the interaction-strength 
$g=1.2$ and an elliptic transverse confinement with 
ellipticity $\epsilon =0.4$. The shape of the bright soliton 
strongly depends on the ellipticity $\epsilon$ of the transverse 
potential and the interaction strength $g$. 
By varying $\epsilon$ and $g$ 
the bright soliton can be spherical-shaped, cigar-shaped, 
disk-shaped, but also fully triaxial. 

Our numerical investigation shows that 
under the condition $\epsilon \geq 0$ the width 
$\sigma_x$ of the soliton along the $x$ axis is always 
close to $1$ (in units of $a_{\bot}$). 
The width $\sigma_y$ of the soliton along the $y$ axis is equal 
to $\sigma_x$ only for $\epsilon=0$; moreover $\sigma_y$ becomes 
extremely large as $\epsilon \to 1$. Obviously, 
with $\epsilon <0$ the behaviors of $\sigma_x$ and $\sigma_y$ 
are interchanged. The width $\sigma_z$ along the $z$ axis 
is instead controlled by the interaction strength $g$: 
a small $g$ implies a very large $\sigma_z$, while 
when $g$ is sufficiently large the width $\sigma_z$ is around $1$. 
In addition it exists a critical strength $g_c$ 
above which there is no solution, i.e. the  
wave function of the metastable soliton collapses to the zero-size 
ground-state of infinite negative energy. 

\begin{figure}
\centering
\includegraphics[width=8.6cm,clip=]{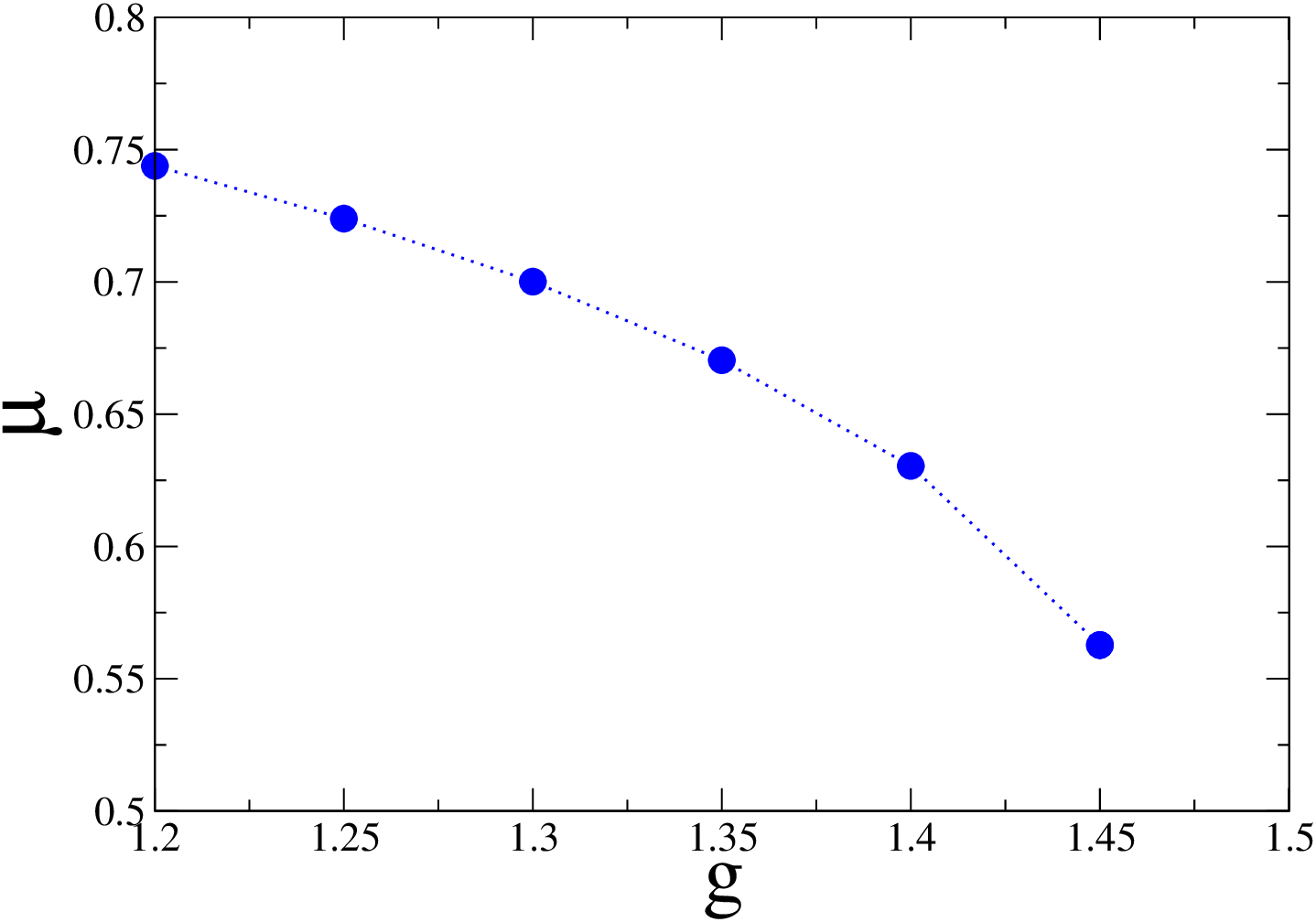}
\caption{Chemical potential $\mu$ 
as a function of the interaction strength $g$ 
obtained with the 3D GPE. $\epsilon=0.8$ is the ellipticity 
of the elliptic transverse harmonic confinement.} 
\label{fig3}
\end{figure}

In Fig. \ref{fig2} we plot this critical strength $g_c$ 
as a function of the ellipticity $\epsilon$ of the 
transverse trap. Our numerical results based on the integration 
of the 3D GPE are displayed as filled circles. 
The figure shows that when the trap is perfectly symmetric ($\epsilon=0$) 
the critical strength $g_c$ reaches its minimum value, 
$g_c = 1.35$. Instead, as $|\epsilon| \to 1^{-}$ 
the critical strength has its maximum value given by 
$g_c=1.66$. We notice that when $|\epsilon| \to 1^{-}$, the frequency 
of confinement along one of the two transverse 
directions goes to zero, but only at $\epsilon=\pm 1$ 
the triaxial bright soliton becomes unbounded. 

\begin{figure}
\centering
\includegraphics[width=8.6cm,clip=]{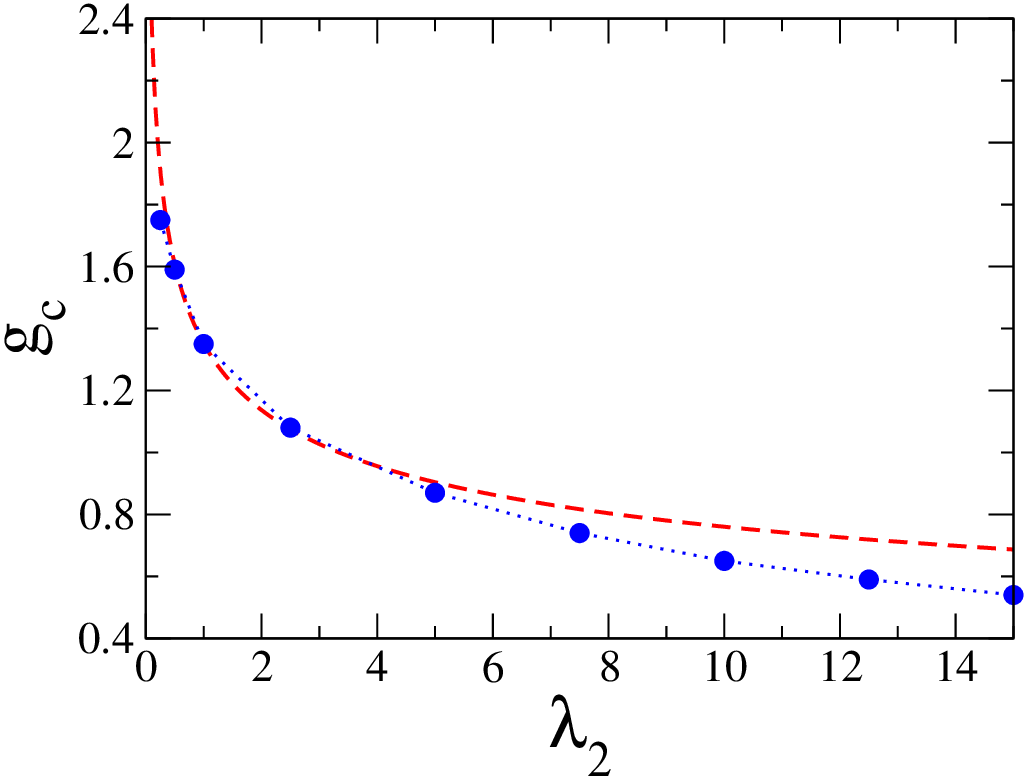}
\caption{Critical strength $g_c$ 
as a function of the scaled frequency $\lambda_2$ with $\lambda_1=1$. 
The interaction strength is $g=2N|a_s|/a_{\bot}$. 
Filled circles: numerical results 
obtained with the 3D GPE. Dashed line: prediction 
of Eq. (\ref{brazil}).} 
\label{fig4}
\end{figure}

It is interesting to compare our results with previous predictions 
based on numerical calculations and scaling \cite{gammal}. 
According to these predictions \cite{gammal} the critical strength $g_c$ 
is simply given by the formula 
\beq 
g_c = {1.352\over (\lambda_1 \lambda_2)^{1/4}} \; . 
\label{brazil}
\eeq 
In Fig. \ref{fig2} the dashed line is obtained with 
Eq. (\ref{brazil}) and the scaled frequencies 
given by Eq. (\ref{ellisse}). The figure shows that there 
are relevant differences between our numerical results (filled circles) 
and Eq. (\ref{brazil}) for $|\epsilon|$ close to $1$. In fact, 
Eq. (\ref{brazil}) implies that $g_c\to+\infty$ for 
$|\epsilon|\to 1^{-}$. Actually, Eq. (\ref{brazil}) is based 
on the hypothesis of a attractive BEC with triaxial harmonic confinement 
of frequencies $\lambda_1$, $\lambda_2$ and $\lambda_3$ 
under the conditions $\lambda_1,\lambda_2\gg \lambda_3$ 
and $\lambda_3\to 0$ \cite{gammal}, but these conditions on the 
harmonic frequencies break down for $|\epsilon|$ close to $1$. 
As previously, discussed, 
our numerical results suggest instead a finite value of $g_c$ 
for $|\epsilon|\to 1^{-}$, as confirmed also by a fully 
Gaussian variational approach \cite{SalaLP}. 

An important issue is the dynamical stability of the 
triaxial bright solitons we have found. According to the 
Vachitov-Kolokolov criterion \cite{vk}, the fundamental solitons 
are stable if they satisfy the condition $d\mu/dg<0$. 
We have verified that up to the collapse this condition is always 
satisfied by our bright solitons. For completeness, 
in Fig. \ref{fig3} we show the calculated chemical potential 
$\mu$ versus the interaction strength $g$ for the triaxial 
bright solitons with ellipticity $\epsilon=0.8$. 

Let us now investigate the general case where $\lambda_1$ and 
$\lambda_2$ are independent. 
Keeping fixed one of the two harmonic frequencies, e.g. $\lambda_1$, 
we may independently tune the other, $\lambda_2$. 
Without loss of generality we fix $\lambda_1=1$. 
We find that the critical strength $g_c$ approaches a 
maximum finite value when the trapping frequency $\lambda_2$ tends 
to zero. This effect is shown in Fig. \ref{fig4}, where we 
plot the critical strength $g_c$ as a function of $\lambda_2$  
with $\lambda_1=1$. Instead, for large values of $\lambda_2$ 
the critical strength $g_c$ becomes smaller. By using a Gaussian 
variational approach \cite{SalaLP} we have indeed verified that $g_c\to 0$ 
as $\lambda_2\to+\infty$. For the sake of completeness, 
in Fig. 4 we have included also the prediction of Eq. (\ref{brazil}) 
with $\lambda_1=1$. Remarkably, there are deviations not only 
for small values of $\lambda_2$ but also for large values 
of $\lambda_2$. 

In conclusion, in this work we have investigated the collapse of 
triaxial bright solitons in 
Bose-condensed atoms under transverse anisotropic harmonic 
confinement by using the 3D Gross-Pitaevskii equation. 
Our predictions on the stability domain of these triaxial bright 
solitons can be useful for future experimental investigations 
with deformed atomic waveguides. 

This work has been partially supported by Fondazione CARIPARO. 
The authors thank Boris Malomed and Flavio Toigo for useful suggestions. 

\section*{Appendix: Numerical method}

The numerical integration of the time dependent GPE, in
Eq.~(\ref{tdgpe}), is obtained by using a finite-difference 
Crank-Nicolson scheme modified with a split operator technique,
adapted to the integration of a Schr\"odinger equation \cite{sala-numerics}. 
This approach has been successfully applied in various problems 
and configurations \cite{8,9}. 

First, we write Eq.~(\ref{tdgpe}) in the form
\beq
i \hbar \frac{\partial}{\partial t} \Psi({\bf r}, t) =
\left( H_{1}({\bf r},t) + H_{2}({\bf r},t) + 
H_{3}({\bf r},t)\right) \Psi({\bf r},t),
\label{gpenum}
\eeq
where
\beq
H_{\alpha} ({\bf r},t) \equiv - \frac{\hbar^{2}}{2m}\frac{ \partial^{2}}
{\partial x_{\alpha}^{2}} + U(x_{\alpha}) 
- \frac{1}{3} g  |\Psi({\bf r},t)|^2 , 
\eeq
with $\alpha = 1,2,3$ and $x_{1}=x$, $x_{2}=y$, $x_{3}=z$. 
Here we have used the fact that the external potential is separable: 
$U(x,y,z)=U(x)+U(y)+U(z)$. 
In this way we split the full Hamiltonian in three sub-Hamiltonians, 
so that at each time we have to write the Laplacian with respect to 
one coordinate only, leading to the solution of a
tridiagonal system, and to huge savings in computer memory.

Equation (\ref{gpenum}) is integrated using the splitted Crank-Nicolson scheme 
\begin{eqnarray}
\Psi ({\bf r}, t + \delta_t)
&=&\frac{1}{1+A_{2}(t)/2}\left(1-A_{1}(t)/2\right) \times \nonumber \\
& &\frac{1}{1+A_{3}(t)/2}\left(1-A_{3}(t)/2\right) \times \nonumber \\
& &\frac{1}{1+A_{1}(t)/2}\left(1-A_{2}(t)/2\right) \Psi ({\bf r} ,t).
\end{eqnarray}
where $\delta_t$ is the integration time step, and $A_{\alpha}(t)
\equiv i \delta_t H_{\alpha}({\bf r},t)/\hbar$. 
The splitting is carried out so that the
commutators are exact up to the order $\delta_t^{2}$ included.
There is obviously a problem with the nonlinear term $g |\Psi ({\bf r},
t)|^{2}$, because we should really use a $\Psi$ somehow averaged over
the time step $\delta_t$, not a $\Psi$ evaluated at the beginning of the
time step. 
To circumvent this problem, we use a predictor-corrector step, 
where each integration step is really done in two times:
going from the time $t$ to the time $t+\delta_t$, the first time we
used $\Psi({\bf r}, t)$ in the nonlinear term, obtaining a
``predicted'' $\tilde \Psi ({\bf r}, t+\delta_t)$; we then repeated the
integration step, starting again from $\Psi ({\bf r}, t)$, but using
$\frac{1}{2} \left (\Psi ({\bf r}, t) + \tilde \Psi ({\bf r},
t+\delta) \right )$ in the nonlinear term. 

In our numerical method the wave function is discretized in the following way 
\beq 
\Psi({\bf r},t) = \Psi(x^i,y^j,z^k,t^s)
\eeq
where $x^i=x_0+i\ \delta_x$, $y^j=y_0+j\ \delta_y$, $z^k=z_0+k\ \delta_z$, 
and $t^s=s\ \delta_t$, with $i,j,k,s$ integer numbers. 
Second derivatives are approximated by finite-difference formulas. 
For instance, along the $x$ axis we use 
\beq 
\frac{ \partial^{2}}{\partial x^2} \Psi(x^i,y^j,z^k,t^s) = 
{\Psi(x^{i+1},y^j,z^k,t^s) - 2 \Psi(x^i,y^j,z^k,t^s) 
+ \Psi(x^{i-1},y^j,z^k,t^s) \over \delta_x^2} \; . 
\eeq

\begin{figure}
\centering
\includegraphics[width=8.6cm,clip=]{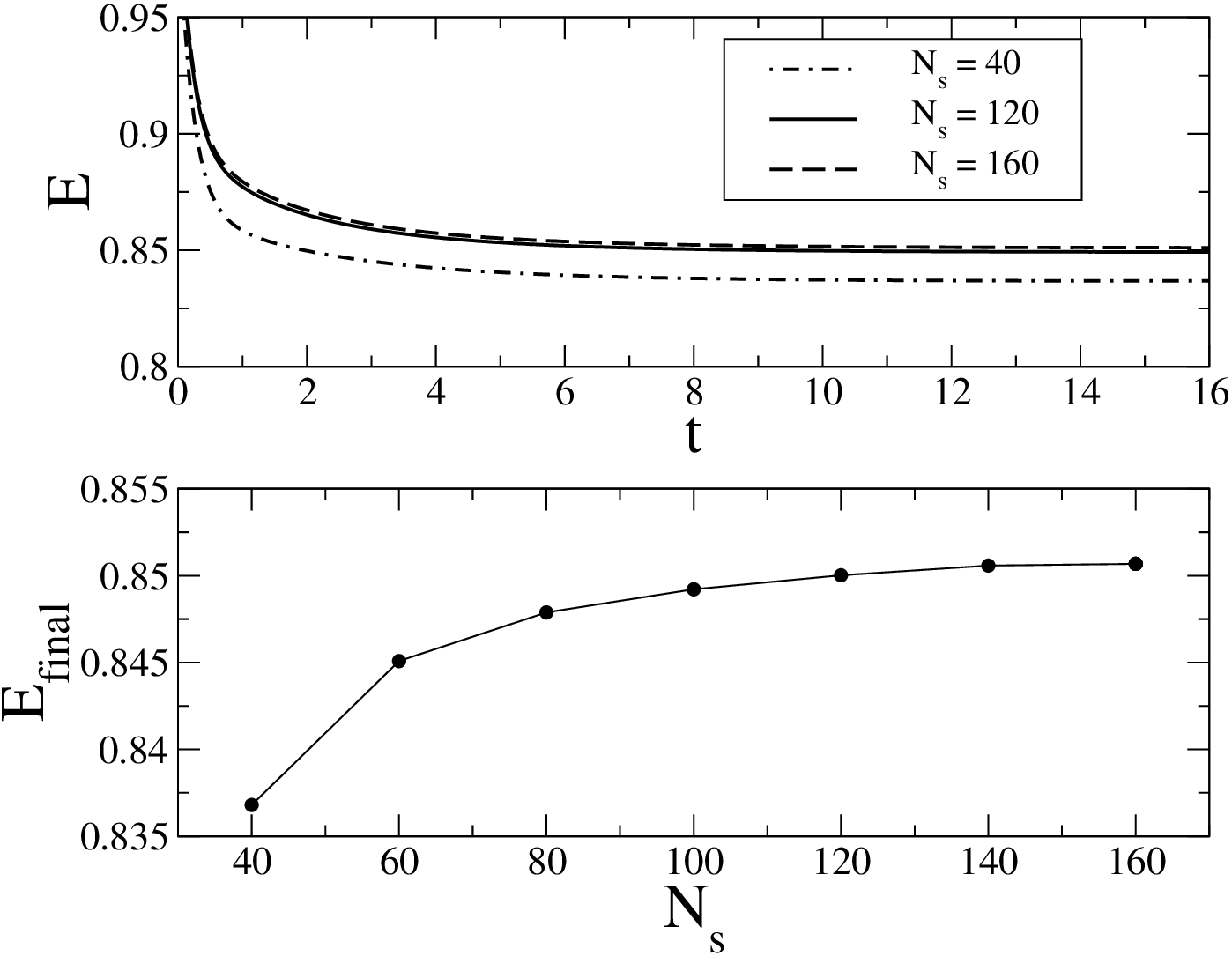}
\caption{Triaxial bright soliton with $g=1.2$ and $\epsilon=0.8$. 
Upper panel: energy $E$ of the 
bright soliton as a function of the imaginary time $t$ 
for 3 values of $N_s$. The spatial mesh has $N_s\times N_s \times 
N_s$ points. Lower panel: asymptotic energy $E_{final}$ of the bright soliton 
as a function of $N_s$.} 
\label{fig5}
\end{figure}

In the upper panel of Fig. \ref{fig5} we show the typical behavior of the 
energy $E$ of the triaxial bright soliton as a function of the 
imaginary time $t$. We use a triaxial Gaussian as initial trial 
wave function $\Psi({\bf r},t=0)=\Psi_{initial}({\bf r})$, 
normalizing to one the norm of $\Psi({\bf r},t)$ at each time step. 
The energy $E$ of the system, given by 
\beq 
E = \int  \Psi^*({\bf r},t) 
\left[ -{\frac{1}{2}}\nabla^{2} + {1\over 2} \left(
\lambda_1^2 x^2 +\lambda_2^2 y^2 \right) - {1\over 2} (2\pi g) 
|\Psi({\bf r},t)|^{2}\right] \Psi({\bf r},t) \ d^3{\bf r} \; , 
\eeq
decreases during the (imaginary-)time evolution 
and eventually reaches its asymptotic value $E_{final}$. 
We find that the asymptotic value $E_{final}$ depends of the number 
$N_s \times N_s \times N_s$ of points in the spatial mesh. 
Nevertheless, as shown in the lower panel of Fig. \ref{fig5}, 
for sufficiently large values of $N_s$ 
the energy $E_{final}$ saturates to the exact value. 
As a final check, we have verified that smaller values of 
the time step $\delta_t$, with respect to the one we use 
($\delta_t=0.05$), do not modify the final results within the third digit. 
Moreover, we have checked that the final energy $E_{final}$ 
and the final wave function $\Psi_{final}({\bf r})$ 
do not depend on the initial trial wavefunction $\Psi_{initial}({\bf r})$. 
$\Psi_{final}({\bf r})$ is the wave function of 
triaxial bright soliton. In the case of collapse, we find that 
the final energy is $E_{final}=-\infty$ and the final wave function 
is a Dirac delta peak, centered at ${\bf r}=0$. 

A very recent and complete review of numerical methods used 
to solve the Gross-Pitaevskii equation has been written by 
Muruganandam and Adhikari \cite{sadhan-num}. In this paper 
the finite-difference Crank-Nicolson scheme we have used 
is explained with many details.

\end{document}